\documentclass[floatfix,aps,superscriptaddress,prd,twocolumn,amsmath,nofootinbib,preprintnumbers]{revtex4-1}%

\usepackage{amssymb}
\usepackage{amsmath}
\usepackage{graphicx}
\usepackage{makecell}
\usepackage[colorlinks=true, citecolor=blue, linkcolor=blue, urlcolor=blue]{hyperref}

\begin{document}

\title{Hubble constant and sound horizon from the late-time Universe}

\author{Xue Zhang}
\email{Corresponding author: zhangxue@yzu.edu.cn}
\affiliation{Center for Gravitation and Cosmology, College of Physical Science and Technology, \\
	Yangzhou University, Yangzhou 225009, China}
\affiliation{CAS Key Laboratory of Theoretical Physics, Institute of Theoretical Physics, \\
    Chinese Academy of Sciences, Beijing 100190, China}

\author{Qing-Guo Huang}
\email{Corresponding author: huangqg@itp.ac.cn}
\affiliation{CAS Key Laboratory of Theoretical Physics, Institute of Theoretical Physics, \\
    Chinese Academy of Sciences, Beijing 100190, China}
\affiliation{School of Physical Sciences, University of Chinese Academy of Sciences, \\
	No. 19A Yuquan Road, Beijing 100049, China}
\affiliation{School of Fundamental Physics and Mathematical Sciences, \\
    Hangzhou Institute for Advanced Study, UCAS, Hangzhou 310024, China}

\date{\today}

\begin{abstract}
We measure the expansion rate of the recent Universe and the calibration scale of the baryon acoustic oscillation (BAO) from low-redshift data.
BAO relies on the calibration scale, i.e., the sound horizon at the end of drag epoch $r_d$,
which often imposes a prior of the cosmic microwave background (CMB) measurement from the Planck satellite.
In order to make really independent measurements of $H_0$,
we leave $r_d$ completely free and use the BAO data sets combined with the 31 observational $H(z)$ data,
GW170817 and Pantheon sample of Type Ia supernovae.
In $\Lambda$CDM model, we get $H_0=68.63^{+1.75}_{-1.77}$ km s$^{-1}$ Mpc$^{-1}$, $r_d=146.85^{+3.29}_{-3.77}$ Mpc.
For the two model-independent reconstructions of $H(z)$,
we obtain $H_0=68.02\pm1.82$ km s$^{-1}$ Mpc$^{-1}$, $r_d=148.18^{+3.36}_{-3.78}$ Mpc in the cubic expansion,
and $H_0=68.58\pm1.76$ km s$^{-1}$ Mpc$^{-1}$, $r_d=148.02^{+3.63}_{-3.60}$ Mpc in the polynomial expansion. The values of Hubble constant $H_0$ and sound horizon $r_d$ are consistent with
the estimate derived from the Planck CMB data assuming a flat $\Lambda$CDM model, but $H_0$ is in $2.4\sim2.6$ $\sigma$ tension with SH0ES 2019, respectively.
\end{abstract}

\maketitle

\section{Introduction}

In the past few years, cosmological parameters have been measured with unprecedented precision.
In particular, the cosmic microwave background (CMB) experiments, such as WMAP and Planck, played a key role.
The Planck Collaboration presents the strongest constraints to date on key parameters, such as the Hubble constant $H_0$.
$H_0$ cannot be measured by CMB experiments directly, but can be inferred once the other cosmological parameters are determined by global fitting.
In the $\Lambda$CDM model, Planck finds a lower value of $H_0$ in the first data release \cite{Ade:2013zuv}
and reports the updated results, $H_0=67.27\pm0.60$ km s$^{-1}$ Mpc$^{-1}$, in the final data release \cite{Aghanim:2018eyx}.
The constraint of $H_0$ in CMB measurement relies on the choice of cosmological model.
At present, although the $\Lambda$CDM model is basically successful in fitting available cosmological data,
it is still challenged by some compatibility tests at low and high redshifts.
Recently, the discrepancy in the Hubble constant measured from low- and high-redshift probes has attracted a lot of attention.
In particular, SH0ES (Supernovae and $H_0$ for the Equation of State) project \cite{Riess:2011yx} constructed a local distance ladder approach
from the Cepheids to measure $H_0$.
The local measurement of $H_0$ is model independent as it does not depend on cosmological assumptions.
They improve the accuracy of $H_0$ and publish the updated results as $H_0=74.03\pm1.42$ km s$^{-1}$ Mpc$^{-1}$ \cite{Riess:2019cxk},
which increases the tension with the final result of Planck to $4.4\sigma$.

In the absence of systematic errors in both measurements,
the model-dependent CMB measurement should be consistent with the model-independent local measurements if the standard cosmological model is correct.
The tension could provide evidence of physics beyond the standard model.
With clear motivation, extensive research has been done on extended models beyond the standard model to alleviate inconsistencies between data sets. For example, see Refs. \cite{DiValentino:2016hlg,Qing-Guo:2016ykt,Zhao:2017cud,Sola:2017znb,Miao:2018zpw,Xu:2016ddc,Guo:2018ans,Yang:2018qmz,Poulin:2018cxd,Ryan:2019uor,Li:2019ypi,Vagnozzi:2019ezj,Liu:2019awo,Ding:2019mmw,DiValentino:2020hov}.
On the other hand, a growing number of other measurements independently provide measurements of the Hubble constant.
The H0LiCOW Collaboration \cite{Suyu:2016qxx} presents another independent approach to measure $H_0$ by the time delay from lensing.
In a flat $\Lambda$CDM cosmology, they provide a latest value $H_0=73.3^{+1.7}_{-1.8}$ km s$^{-1}$ Mpc$^{-1}$ (2.4\% precision) \cite{Wong:2019kwg}.
It is consistent with the local measurement of $H_0$ by the distance ladder, but in 3.2$\sigma$ tension with respect to the CMB data from Planck satellite. This method is independent of both the distance ladder and other cosmological probes.
In addition, the Advanced LIGO and Virgo report a gravitational-wave measurement of the Hubble constant $H_0=70^{+12}_{-8}$ km s$^{-1}$ Mpc$^{-1}$
using the gravitational-wave signal from the merger of a binary neutron-star system \cite{Abbott:2017xzu}.
The red giant branch method provides one of the most accurate means of measuring the distances to nearby galaxies.
Recently, using the revised measurement, Ref. \cite{Freedman:2020dne} reported $H_0=69.6\pm0.8$ km s$^{-1}$ Mpc$^{-1}$.

The baryon acoustic oscillation (BAO) surveys provide measurements of three types $D_A(z)/r_d$, $D_V(z)/r_d$ and $H(z)r_d$,
where $r_d$ is the comoving size of sound horizon at the end of the baryon drag epoch \cite{Eisenstein:1997ik,Hu:1996vq}.
The Hubble constant $H_0$ and sound horizon $r_d$ are closely related and link the late-time and early time cosmology.
If we measure $H_0$ using the BAO data, an independent distance calibration is required.
In other words, $r_d$ is the standard ruler which calibrates the distance scale measurements of BAO.
In general, $r_d$ relies on the physical properties of the early universe, which can be constrained by precise CMB observations.
The CMB measurement relies on the assumption of a $\Lambda$CDM model to constrain the cosmological parameters.
In most all of BAO measurements $r_d$ often be imposed a Gaussian prior to $r_d$ from CMB. In this sense, the constraint on the Hubble constant by using BAO data, for example, Ref. \cite{Zhang:2019cww}, is not completely independent on the CMB data.
Instead of early time physical calibration of $r_d $, an alternative approach is to combine BAO measurements with other low-redshift observations.

Planck public available MCMC chains give $r_d=147.05\pm0.30$ Mpc in $\Lambda$CDM model.
This is a model-dependent theoretical expectation determined from the CMB measurement.
Assuming the cold dark matter model with a cosmological constant,
Ref. \cite{Heavens:2014rja} took the sound horizon at radiation drag as a ruler,
determined $r_d=142.8\pm3.7$ Mpc by adding the clocks and the local $H_0$ measurement to the SNe and BAO.
They found excellent agreement with the derived quantity of the sound horizon deduced from Planck data.
In the spline models for the expansion history $H(z)$,
Bernal et al. obtained $r_d=136.8\pm4.0$ Mpc and $r_d=133.0\pm4.7$ Mpc when $\Omega_k$ was left as a free parameter
from the BAO, SNe Ia, and local measurement without CMB-derived $r_d$ prior \cite{Bernal:2016gxb}.
Combining the data sets from clocks, SNe, BAO, and local measurement of $H_0$,
Verde et al. found $r_d=143.9\pm3.1$ Mpc with a flat curvature \cite{Verde:2016ccp}.
Then, using BAO measurements and SNe Ia, calibrated with time delay from H0LiCOW,
Aylor et al. inferred the sound horizon $r_d=139.3^{+4.8}_{-4.4}$ Mpc in $\Lambda$CDM model \cite{Aylor:2018drw}.
Using the inverse distance ladder method,
Dark Energy Survey Collaboration found $r_d=145.2\pm18.5$ Mpc from SNe Ia and BAO measurements \cite{Macaulay:2018fxi}.
In their analysis, they adopted a prior on $r_d$ taken from the Planck 2018.
Using the supernovae Ia and BAO measurements combined with $H_0$ from H0LiCOW,
Ref. \cite{Wojtak:2019tkz} provides the sound horizon at recombination $r_d=137.0\pm4.5$ Mpc in the polynomial expansion of $H(z)$.
See Refs. \cite{LHuillier:2016mtc,Shafieloo:2018gin,Camarena:2019rmj,Nunes:2020hzy}
for more papers about the sound horizon.
This apparent discrepancy comes from fitting the BAO measurements with or without a prior of CMB from Planck.
Comparing the sound horizon obtained from the low-redshift data with the value derived from Planck
may give us a better understanding of the discordance between the data sets or reveal new physics beyond the standard model.

In order to solve the discrepancy of $H_0$ and $r_d$ from early and late universe,
using the recent low-redshift data to constrain the sound horizon of early universe is the main motivation of this paper.
In our analysis, we consider the $\Lambda$CDM model and two model-independent reconstructions of $H(z)$.
Without any assumption about the early time physics, we set the standard ruler $r_d$ of BAO as a free parameter.
Combining BAO with observational $H(z)$ data, gravitational wave, and SN Ia measurement,
we measure the Hubble constant $H_0$ and sound horizon $r_d$ regardless of the early time physics.
In Sec. \ref{sec:hz} we introduce the reconstruction of $H(z)$.
The data sets and methodology used in this paper are shown in Sec. \ref{sec:data}.
In Sec. \ref{sec:result} we present the results of sound horizon without assuming any early time physics.
We summarize the conclusions in the last section.


\section{The reconstruction of $H(z)$}
\label{sec:hz}

We will perform our analyses with following three different forms of $H(z)$.

First, in flat $\Lambda$CDM model, the Hubble parameter can be expressed as
\begin{equation}
H(z)=H_0\sqrt{\Omega_\mathrm{m} (1+z)^3 + \Omega_\Lambda} ,
\end{equation}
where $\Omega_\Lambda=1-\Omega_\mathrm{m}$.

Second,
in order to avoid working within a specific cosmological model,
we try to reconstruct $H(z)$ in the a model-independent way.
The Hubble parameter is expressed as a cubic expansion of scale factor $(1-a)$,
\begin{equation}
H(z)=H_0\left[1 + h_1(1-a) + h_2(1-a)^2 + h_3(1-a)^3\right].
\end{equation}
We can easily determine $H_0$ from the corresponding reconstructed $H(z)$ ranges.

The third one is a polynomal expansion of $H(z)$.
We follow \cite{Zhang:2016urt}
and Taylor expand the scale factor with respect to cosmological time.
Then the Hubble parameters $H(t)$, deceleration parameters $q(t)$,
jerk parameters $j(t)$ and snap parameters $s(t)$ are defined as
\begin{align}
  H(t) =& +\frac{1}{a}\frac{da}{dt}, \\
  q(t) =& -\frac{1}{a}\frac{d^2a}{dt^2} \left[\frac{1}{a}\frac{da}{dt}\right]^{-2}, \\
  j(t) =& +\frac{1}{a}\frac{d^3a}{dt^3} \left[\frac{1}{a}\frac{da}{dt}\right]^{-3}, \\
  s(t) =& +\frac{1}{a}\frac{d^4a}{dt^4} \left[\frac{1}{a}\frac{da}{dt}\right]^{-4}.
\end{align}
Using these parameters, the Hubble parameter can be parametrized as a polynomial expansion
\begin{align}\nonumber
 H(z)=& H_0+{\frac{dH}{dz}}\Big|_{z=0}z
        +\frac{1}{2!}{\frac{d^2H}{dz^2}}\Big|_{z=0}z^2
        +\frac{1}{3!}{\frac{d^3H}{dz^3}}\Big|_{z=0}z^3\\\nonumber
        &+\frac{1}{4!}{\frac{d^4H}{dz^4}}\Big|_{z=0}z^4 + \cdots\\\nonumber
     =& H_0 \Big[1+(1+q_0)z+\frac{1}{2}(j_0-q_0^2)z^2\\
     &+\frac{1}{6}(3q_0^2+3q_0^3-4q_0j_0-3j_0-s_0)z^3+\mathcal{O}(z^4)\Big]
\end{align}
where the subscript ``0" indicates the parameters at the present epoch ($z=0$).

\section{Data}
\label{sec:data}

We use the observational data sets including the measurements of the BAO, observational $H(z)$ data (OHD), GW170817 and Pantheon sample.
For BAO measurement, the angular diameter distance $D_A$ and the volume-averaged scale $D_V$ are related to $H(z)$ by
\begin{align}
D_A (z) &= \frac{1}{1+z} \int^z_0 \frac{d z'}{H(z')}, \\
D_V (z) &= \left[ (1+z)^2 D_A^2 (z) \frac{z}{H(z)} \right] ^{1/3}.
\end{align}
The sound horizon is given by
\begin{equation}
r_d=\int^{\infty}_{z_d} \frac{c_s(z)}{H(z)} dz,
\end{equation}
where $c_s(z)$ is the sound speed and $z_d$ is the redshift at the end of drag epoch.
The sound horizon $r_d$ is the standard ruler to calibrate the BAO observations \cite{Heavens:2014rja,Verde:2016ccp}.
It is often imposed a prior of the CMB measurement from Planck satellite.
In this paper, we remove the prior $r_d$ from Planck and set $r_d$ as a free sampling parameter.

We use the constraints on BAO from the following galaxy surveys:
the 6dF Galaxy Survey \cite{Beutler:2011hx}, the SDSS DR7 Main Galaxy sample \cite{Ross:2014qpa},
the BOSS DR12 \cite{Wang:2016wjr}, and eBOSS DR14 quasar \cite{Ata:2017dya}.
We also include eBOSS DR14 Ly$\alpha$ \cite{Agathe:2019vsu}.
The data sets are listed in Table \ref{tab:BAO}.
\begin{table}[!htbp]
\centering
\caption{BAO data measurements included in our analysis.
$D_A$, $D_V$, and $r_d$ are in units of Mpc, while $H(z)$ is in units of km s$^{-1}$ Mpc$^{-1}$.}
\label{tab:BAO}
\begin{tabular}{ccccc}
  \hline
  \hline
   $z_\mathrm{eff}$ & Measurement & Constraint & Reference \\
   \hline
   0.106 & $r_d/D_V$ & $0.336 \pm 0.015$ & \cite{Beutler:2011hx} \\
   0.15 & $D_V/r_d$ & $4.47 \pm 0.17$ & \cite{Ross:2014qpa} \\
   0.31 & $D_A/r_d$ & $6.29 \pm 0.14$ & \cite{Wang:2016wjr} \\
   0.36 & $D_A/r_d$ & $7.09 \pm 0.16$ & \cite{Wang:2016wjr} \\
   0.40 & $D_A/r_d$ & $7.70 \pm 0.16$ & \cite{Wang:2016wjr} \\
   0.44 & $D_A/r_d$ & $8.20 \pm 0.13$ & \cite{Wang:2016wjr} \\
   0.48 & $D_A/r_d$ & $8.64 \pm 0.11$ & \cite{Wang:2016wjr} \\
   0.52 & $D_A/r_d$ & $8.90 \pm 0.12$ & \cite{Wang:2016wjr} \\
   0.56 & $D_A/r_d$ & $9.16 \pm 0.14$ & \cite{Wang:2016wjr} \\
   0.59 & $D_A/r_d$ & $9.45 \pm 0.17$ & \cite{Wang:2016wjr} \\
   0.64 & $D_A/r_d$ & $9.62 \pm 0.22$ & \cite{Wang:2016wjr} \\
   0.31 & $H*r_d$ & $11550 \pm 700$ & \cite{Wang:2016wjr} \\
   0.36 & $H*r_d$ & $11810 \pm 500$ & \cite{Wang:2016wjr} \\
   0.40 & $H*r_d$ & $12120 \pm 300$ & \cite{Wang:2016wjr} \\
   0.44 & $H*r_d$ & $12530 \pm 270$ & \cite{Wang:2016wjr} \\
   0.48 & $H*r_d$ & $12970 \pm 300$ & \cite{Wang:2016wjr} \\
   0.52 & $H*r_d$ & $13940 \pm 390$ & \cite{Wang:2016wjr} \\
   0.56 & $H*r_d$ & $13790 \pm 340$ & \cite{Wang:2016wjr} \\
   0.59 & $H*r_d$ & $14550 \pm 470$ & \cite{Wang:2016wjr} \\
   0.64 & $H*r_d$ & $14600 \pm 440$ & \cite{Wang:2016wjr} \\
   1.52 & $D_V/r_d$ & $26.00 \pm 0.99$ & \cite{Ata:2017dya}\\
   2.34 & $D_A/r_d$ & $11.20\pm0.56$ & \cite{Agathe:2019vsu} \\
  \hline
\end{tabular}
\end{table}


In the current analysis, we do not make use of the OHD extracted from the measurement of BAO.
We only consider the OHD from differential age method.
The differential age method is proposed in Ref. \cite{Jimenez:2001gg}.
It can be used to measure the expansion rate of the Universe.
The quantity measured in differential age method is directly related to the Hubble parameter,
\begin{equation}
H(z)=-\frac{1}{(1+z)}\frac{dz}{dt}.
\end{equation}
This method can be used to determine Hubble constant $H_0$.
Table \ref{tab:ohd} shows an updated compilation of OHD accumulating a total of 31 points given by differential age method \cite{Magana:2017nfs}.
\begin{table}
    \centering
    \caption{The 31 observational $H(z)$ data obtained by the differential age method.}
    \label{tab:ohd}
    \begin{tabular}{l c c }
        \hline
        \hline 
        $z$ & $H(z)$ & Reference \\
        \hline
        0.09  & 69 $\pm$ 12 & \cite{Jimenez:2003iv} \\
        \hline
        0.17 & 83 $\pm$ 8  & \cite{Simon:2004tf} \\
        0.27 & 77 $\pm$ 14 & \cite{Simon:2004tf}\\
        0.4  & 95 $\pm$ 17 & \cite{Simon:2004tf}\\
        0.9  & 117 $\pm$ 23 & \cite{Simon:2004tf}\\
        1.3  & 168 $\pm$ 17 & \cite{Simon:2004tf}\\
        1.43 & 177 $\pm$ 18 & \cite{Simon:2004tf}\\
        1.53 & 140 $\pm$ 14 & \cite{Simon:2004tf}\\
        1.75 & 202 $\pm$ 40 & \cite{Simon:2004tf}\\
        0.48 & 97 $\pm$ 62 & \cite{Stern:2009ep} \\
        0.88 & 90 $\pm$ 40 & \cite{Stern:2009ep}\\
        0.1791 & 75 $\pm$ 4  & \cite{Moresco:2012jh}  \\
        0.1993 & 75 $\pm$ 5  & \cite{Moresco:2012jh}\\
        0.3519 & 83 $\pm$ 14 & \cite{Moresco:2012jh}\\
        0.5929 & 104 $\pm$ 13 & \cite{Moresco:2012jh}\\
        0.6797 & 92 $\pm$ 8  & \cite{Moresco:2012jh}\\
        0.7812 & 105 $\pm$ 12 & \cite{Moresco:2012jh}\\
        0.8754 & 125 $\pm$ 17 & \cite{Moresco:2012jh}\\
        1.037  & 154 $\pm$ 20 & \cite{Moresco:2012jh}\\
        0.07 & 69.0 $\pm$ 19.6 & \cite{Zhang:2012mp} \\
        0.12 & 68.6 $\pm$ 26.2 & \cite{Zhang:2012mp}\\
        0.20 & 72.9 $\pm$ 29.6 & \cite{Zhang:2012mp}\\
        0.28 & 88.8 $\pm$ 36.6 & \cite{Zhang:2012mp}\\
        1.363  & 160 $\pm$ 33.6  & \cite{Moresco:2015cya} \\
        1.965  & 186.5 $\pm$ 50.4  & \cite{Moresco:2015cya}\\
        0.3802 & 83 $\pm$ 13.5 & \cite{Moresco:2016mzx} \\
        0.4004 & 77 $\pm$ 10.2 & \cite{Moresco:2016mzx}\\
        0.4247 & 87.1 $\pm$ 11.2 & \cite{Moresco:2016mzx}\\
        0.4497 & 92.8 $\pm$ 12.9 & \cite{Moresco:2016mzx}\\
        0.4783 & 80.9 $\pm$ 9 & \cite{Moresco:2016mzx}\\
        0.47 & 89 $\pm$ 23 & \cite{Ratsimbazafy:2017vga} \\
        \hline
        \end{tabular}
\end{table}


Recently, the Advanced LIGO and Virgo detectors observed the gravitational-wave event GW170817
which is a strong signal from the merger of a binary neutron-star system \cite{Abbott:2017xzu}.
The measurement of GW170817 reports
\begin{equation}
H_0=70.0^{+12.0}_{-8.0}~\mathrm{km}~\mathrm{s}^{-1}~\mathrm{Mpc}^{-1}.
\end{equation}

In addition, the Pantheon sample includes 1048 SNe Ia which is the largest confirmed SNe Ia sample \cite{Scolnic:2017caz}.

To perform joint analyses of the data sets,
we explore the cosmological parameter space by a likelihood function $\mathcal{L}$ satisfying $-2 \ln \mathcal{L}=\chi^2$.
We calculate the $\chi^2$ function with the following equation.
\begin{equation}
\chi^2=\chi^2_\mathrm{BAO}+\chi^2_\mathrm{OHD}+\chi^2_\mathrm{GW}+\chi^2_\mathrm{SN}.
\end{equation}
All presented results are computed using the public Monte Carlo Markov Chain public code CosmoMC \cite{Lewis:2002ah}.


\section{Results}
\label{sec:result}

\begin{figure}[!htbp]
\centering
\includegraphics[width=.4\textwidth]{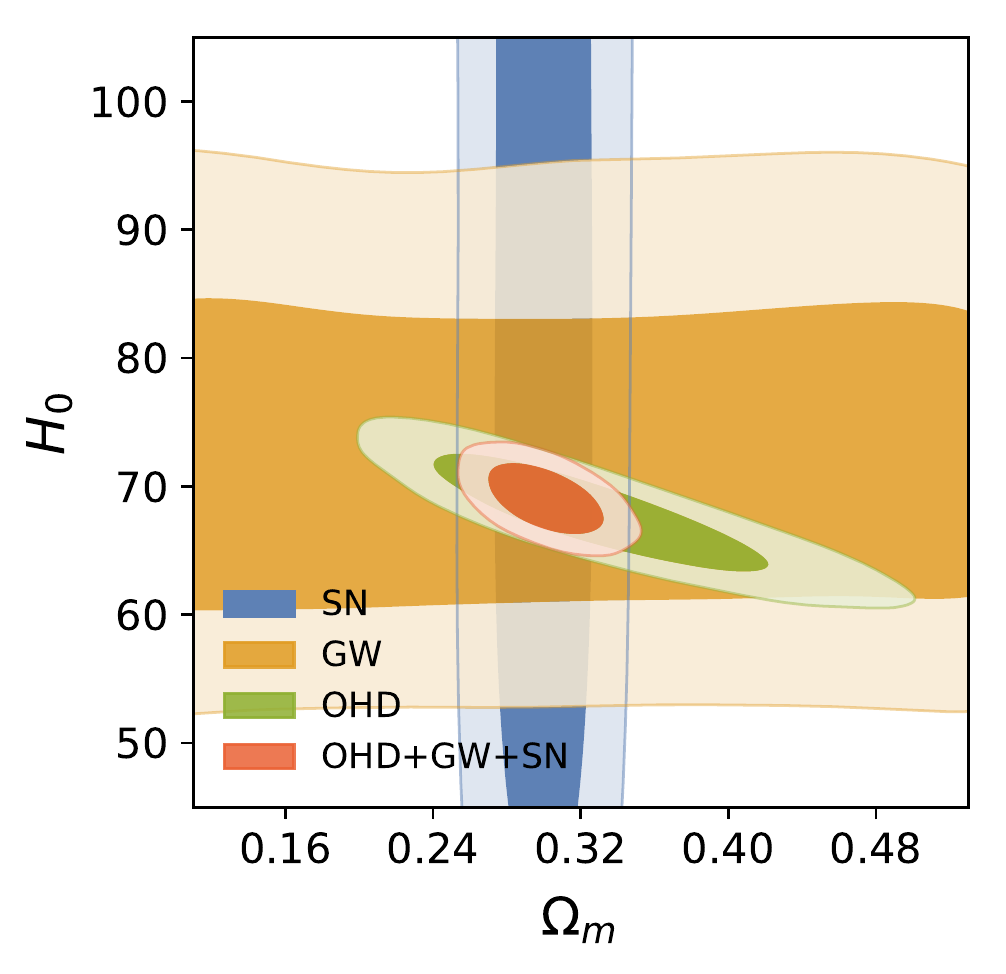}
\includegraphics[width=.4\textwidth]{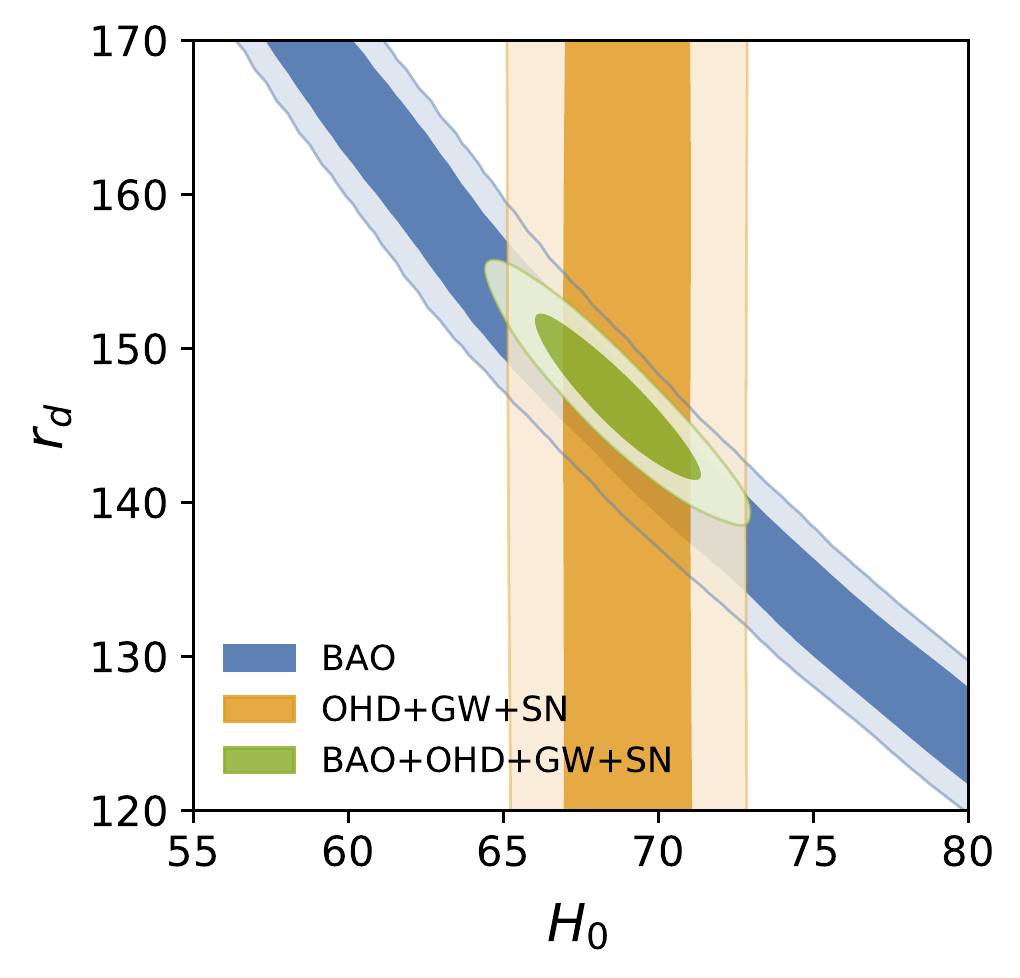}
\caption{The constraints on $\Omega_\mathrm{m}$-$H_0$ and $H_0$-$r_d$ panel in the flat $\Lambda$CDM model using the different data sets.}
\label{fig:lcdm}
\end{figure}

In $\Lambda$CDM model, we set $\Omega_\mathrm{m}$, $H_0$ and $r_d$ as free parameters.
Figure \ref{fig:lcdm} shows our results, including the contours of $\Omega_\mathrm{m}$-$H_0$ and $H_0$-$r_d$
for BAO, OHD, GW, SN, and the joint analyses of the data sets.
The upper figure shows that the SN data sets cannot constrain $H_0$ and the GW data cannot constrain $\Omega_m$.
However, the joint OHD+GW+SN can give a tight constraint of $H_0$.
From the lower figure, we find the OHD+GW+SN data sets cannot constrain $r_d$.
Their joint result gives the constraint of Hubble constant.
The BAO-only data cannot limit the value of $H_0$ or $r_d$.
Combining BAO with the OHD+GW+SN data sets, we get
\begin{align}
H_0&=68.63^{+1.75}_{-1.77}~\mathrm{km}~\mathrm{s}^{-1}~\mathrm{Mpc}^{-1}, \\
r_d&=146.85^{+3.29}_{-3.77}~\mathrm{Mpc}.
\end{align}
The Hubble constant $H_0$ and sound horizon $r_d$ are consistent with the results of Planck 2018.
However, there is a 2.4 $\sigma$ tension on $H_0$ with SH0ES 2019.

\begin{figure*}[!htbp]
\centering
\includegraphics[width=.6\textwidth]{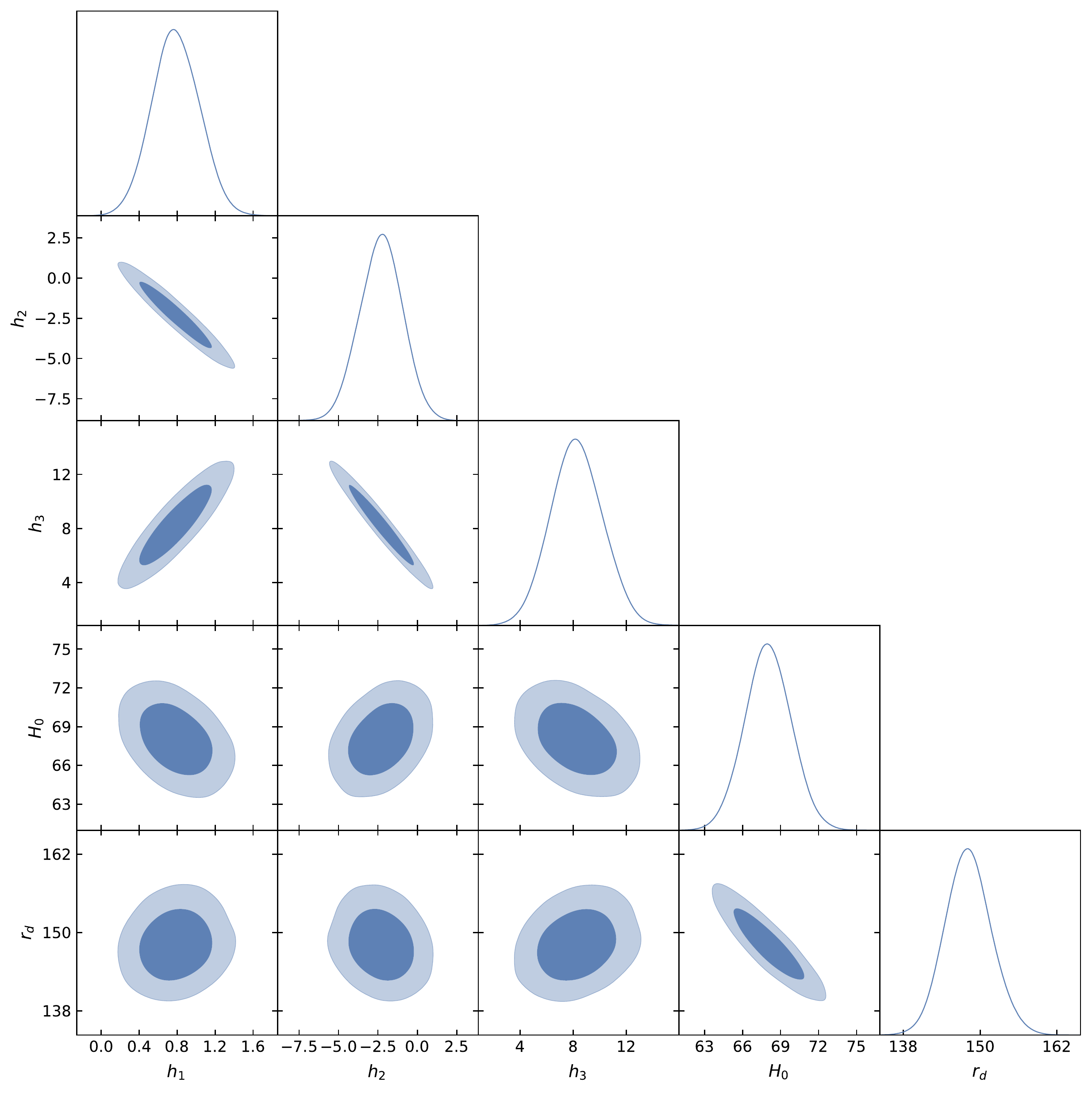}
\caption{The constraints on the parameters in cubic expansion using the BAO+OHD+GW+SN data sets.}
\label{fig:tri_c}
\end{figure*}

\begin{figure*}[!htbp]
\centering
\includegraphics[width=.6\textwidth]{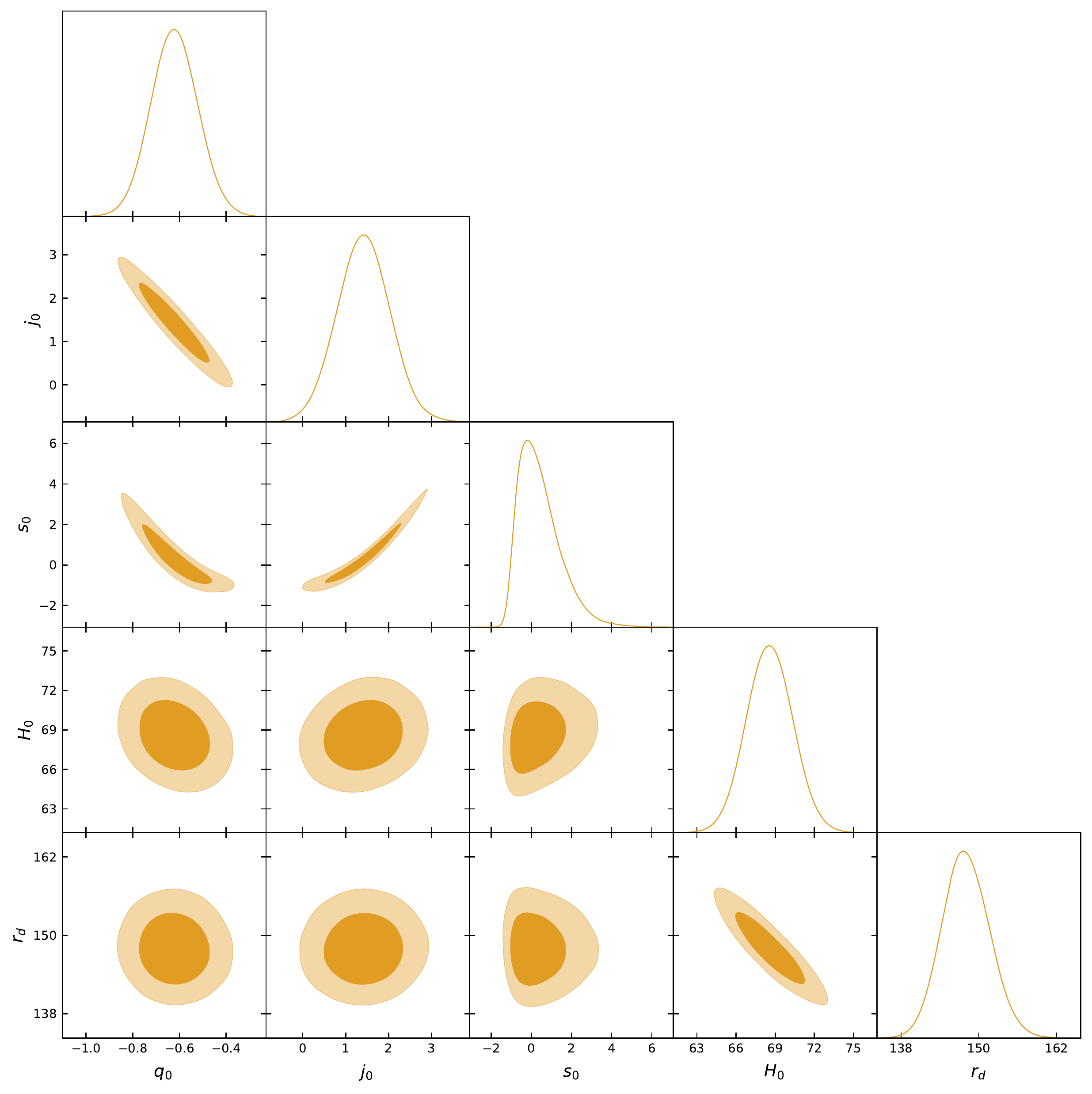}
\caption{The constraints on the parameters in polynomial expansion using the BAO+OHD+GW+SN data sets.}
\label{fig:tri_p}
\end{figure*}

The constraints on the parameters of cubic expansion and polynomial expansion are illustrated in Figs. \ref{fig:tri_c} and \ref{fig:tri_p}.
The blue contours show 68\% and 95\% constraints in cubic expansion using the BAO+OHD+GW+SN data sets without prior on $r_d$.
The orange contours show the constraints cubic expansion using the the BAO+OHD+GW+SN data sets without prior on $r_d$.

\begin{figure}[!htbp]
\centering
\includegraphics[width=.4\textwidth]{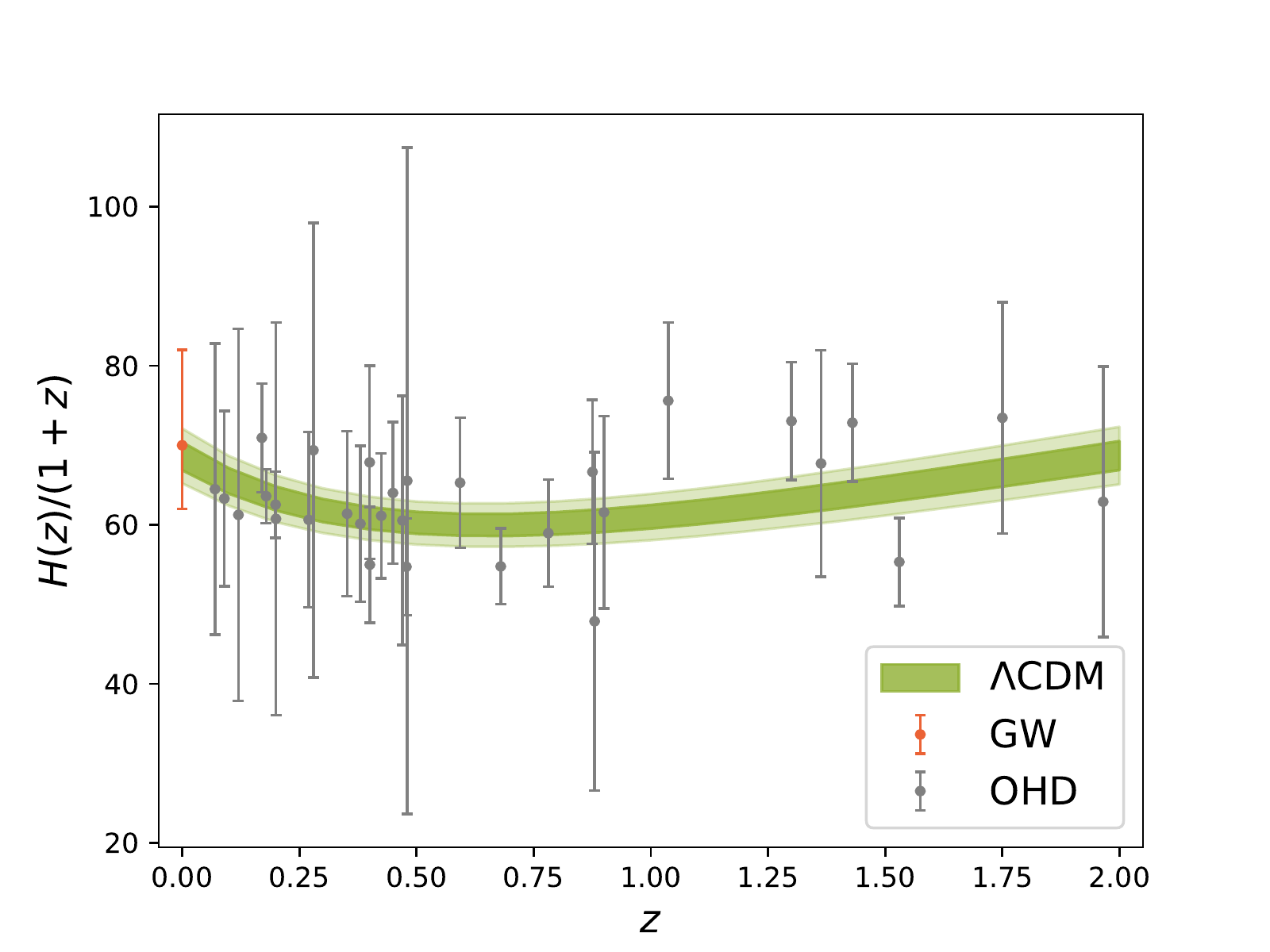} \\
\includegraphics[width=.4\textwidth]{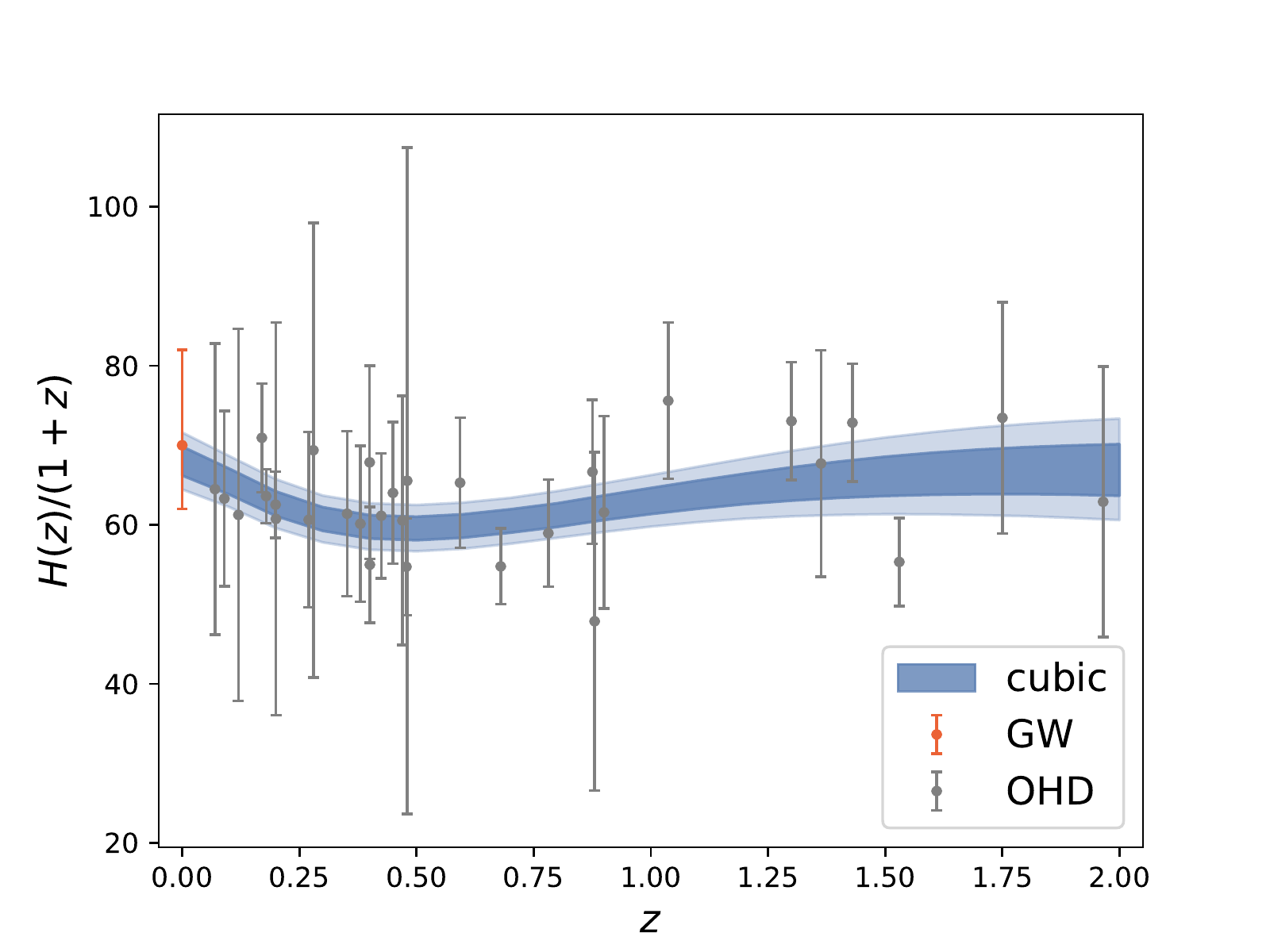}
\includegraphics[width=.4\textwidth]{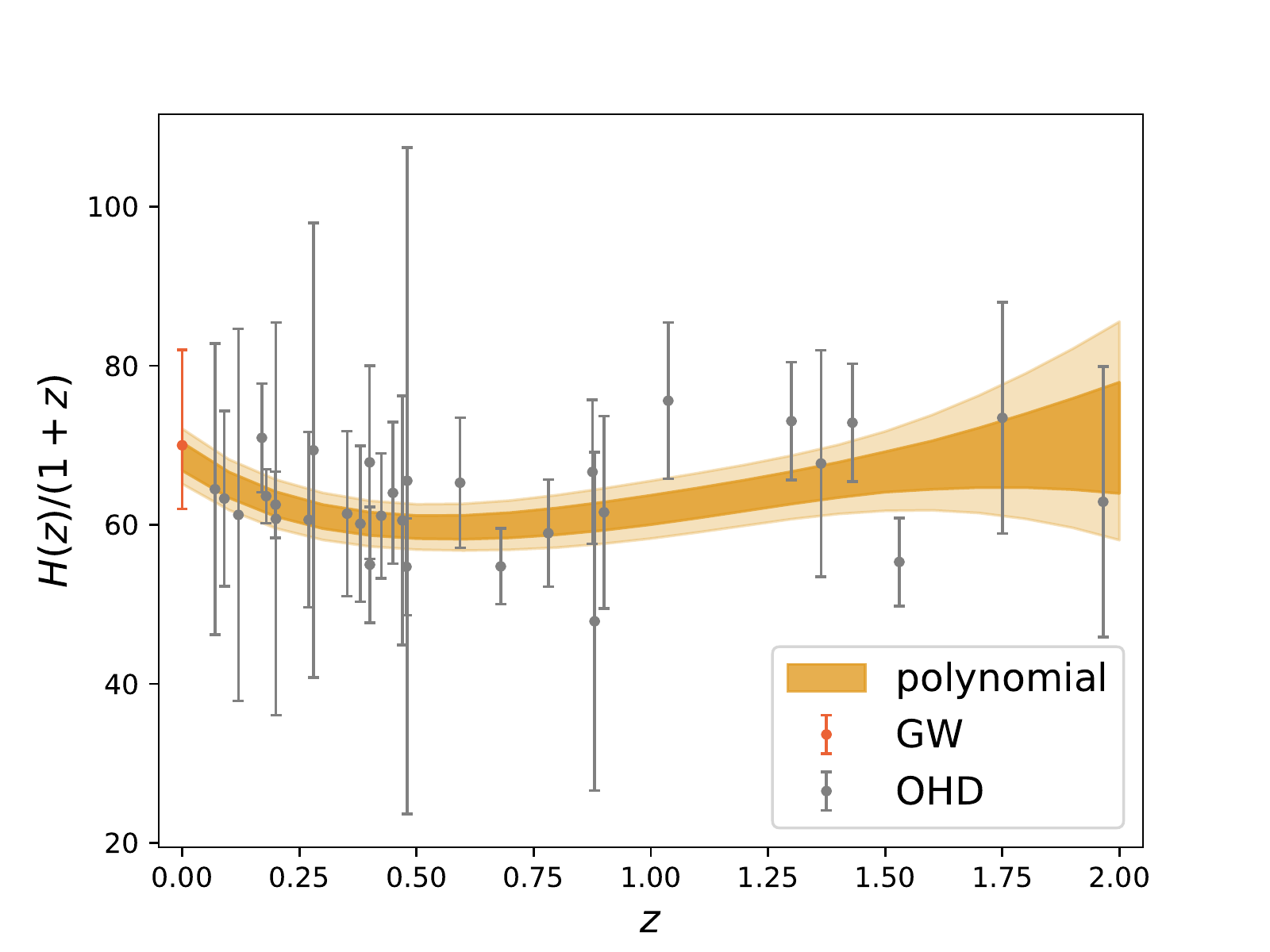}
\caption{Reconstruction of $H(z)$ in (km/s/Mpc) obtained from the BAO+OHD+GW+SN data sets
in $\Lambda$CDM model, cubic expansion and polynomial expansion, respectively.}
\label{fig:errorbar}
\end{figure}

\begin{figure}[!htbp]
\centering
\includegraphics[width=.32\textwidth]{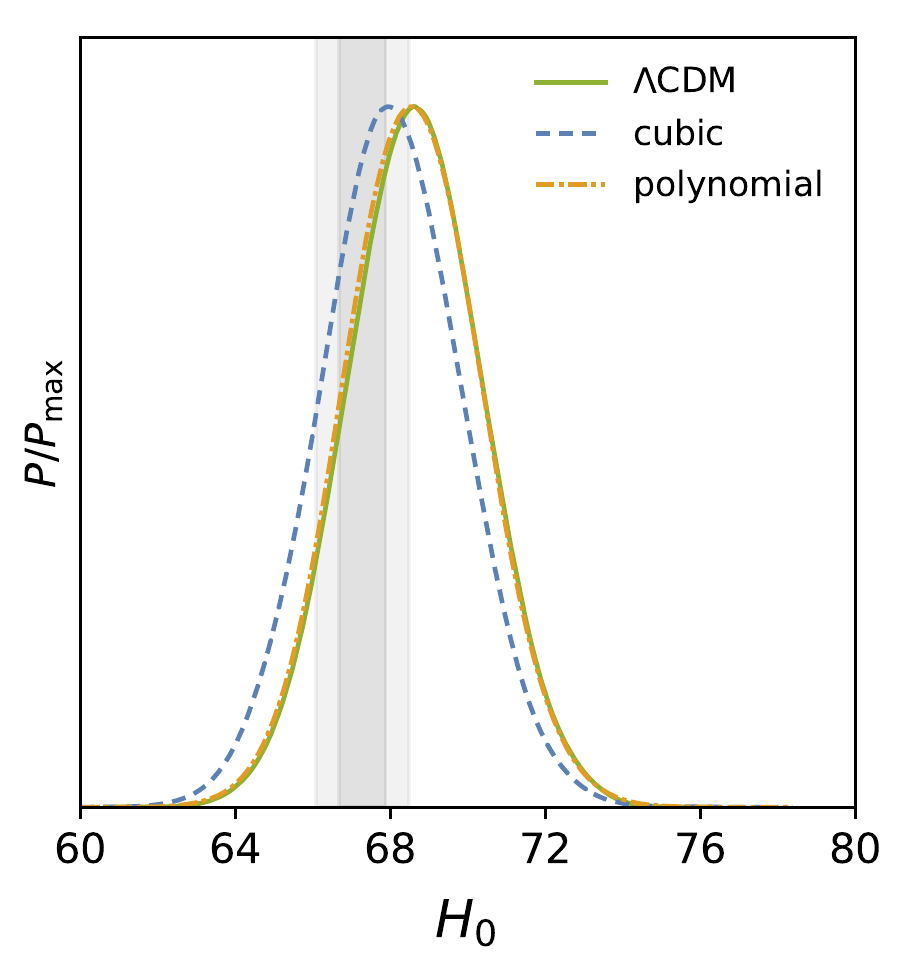}
\includegraphics[width=.32\textwidth]{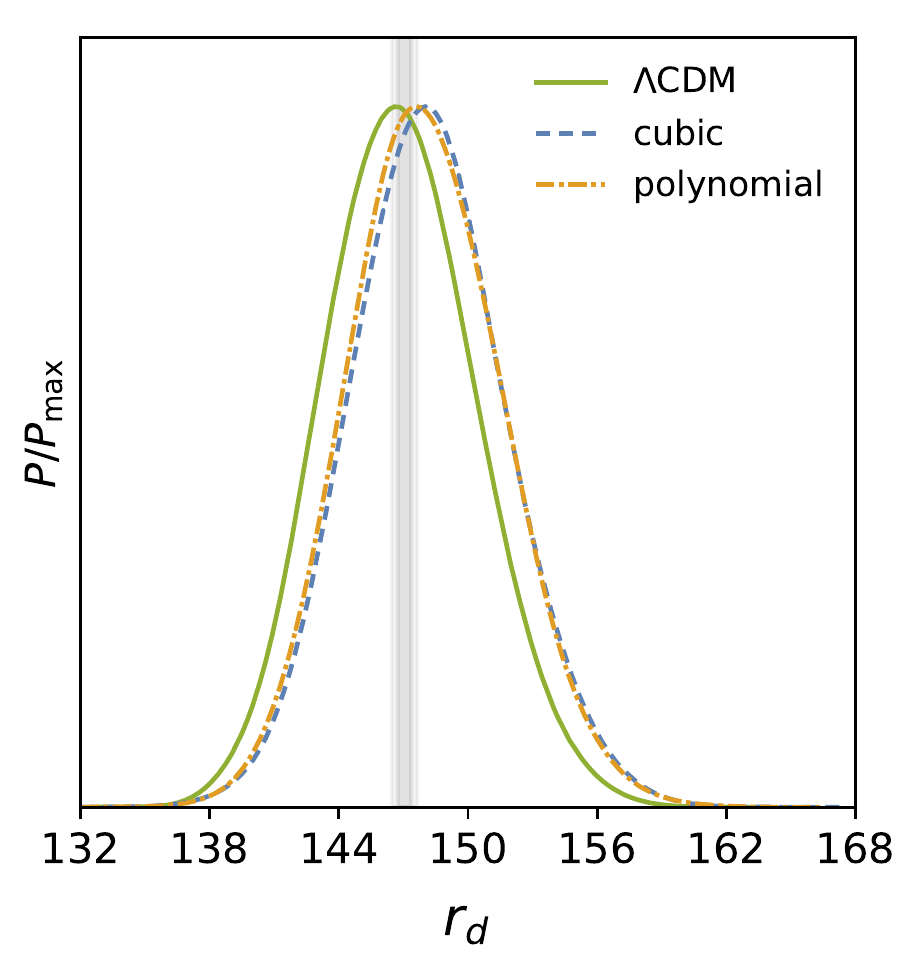}
\caption{The one-dimensional  marginalized distributions of $H_0$ and $r_d$ using the joint datasets BAO+OHD+GW+SN.
For comparison, the gray band represents the 68\% confidence interval of the CMB from Planck 2018.}
\label{fig:h0rd}
\end{figure}

In Fig. \ref{fig:errorbar},
we show the evolutions of expansion histories $H(z)$ on 68\% and 95\% confidence levels using the joint BAO+OHD+GW+SN data sets.
The $\Lambda$CDM model is represented by green region,
the cubic expansion is represented by blue region
and polynomial expansion represented by in orange region.
The OHD data sets are shown in gray and the GW170817 is shown in red.
We display the constraints on $H_0$ and $r_d$ in three different models without CMB prior on $r_d$ in Fig. \ref{fig:h0rd}.
For comparison, the gray bands represent the inferred value by the final result of Planck in $\Lambda$CDM model.

\begin{table}[!htbp]
\centering
\caption{The results in $\Lambda$CDM, cubic expansion and polynomial expansion of $H(z)$ using BAO+OHD+GW+SN.}
\label{tab:result}
\begin{tabular}{ccccc}
  \hline
  \hline
  Parameter & $\Lambda$CDM & Cubic & Polynomial\\
  \hline
  $\Omega_m$ & $0.309\pm0.017$ & $\cdot\cdot\cdot$ & $\cdot\cdot\cdot$ \\
  $h_1$ & $\cdot\cdot\cdot$ & $0.79\pm0.25$  &$\cdot\cdot\cdot$\\
  $h_2$ & $\cdot\cdot\cdot$ & $-2.29^{+1.35}_{-1.37}$  &$\cdot\cdot\cdot$\\
  $h_3$ & $\cdot\cdot\cdot$ & $8.26^{+1.97}_{-1.93}$ & $\cdot\cdot\cdot$\\
  $q_0$ & $\cdot\cdot\cdot$ & $\cdot\cdot\cdot$ & $-0.62\pm0.10$ \\
  $j_0$ & $\cdot\cdot\cdot$ & $\cdot\cdot\cdot$ & $1.42\pm0.61$ \\
  $s_0$ & $\cdot\cdot\cdot$ & $\cdot\cdot\cdot$ & $0.42^{+0.57}_{-1.27}$ \\
  $H_0$ & $68.63^{+1.75}_{-1.77}$ & $68.02\pm1.82$ & $68.58\pm1.76$ \\
  $r_d$ & $146.85^{+3.29}_{-3.77}$ & $148.18^{+3.36}_{-3.78}$ & $148.02^{+3.63}_{-3.60}$ \\
  \hline
\end{tabular}
\end{table}

Table \ref{tab:result} lists the best fit value of parameters inferred from BAO+OHD+GW+SN data sets
in $\Lambda$CDM, cubic expansion and polynomial expansion.
For the cubic expansion on $H(z)$ without a CMB prior on $r_d$, we obtain
\begin{align}
H_0&=68.02\pm1.82~\mathrm{km}~\mathrm{s}^{-1}~\mathrm{Mpc}^{-1}, \\
r_d&=148.18^{+3.36}_{-3.78}~\mathrm{Mpc}.
\end{align}
The Hubble constant $H_0$ and sound horizon $r_d$ are consistent with the results of Planck 2018.
However, there is a 2.6 $\sigma$ tension on the Hubble constant with SH0ES 2019.
For the polynomial expansion on $H(z)$ without a CMB prior on $r_d$, we obtain
\begin{align}
H_0&=68.58\pm1.76~\mathrm{km}~\mathrm{s}^{-1}~\mathrm{Mpc}^{-1}, \\
r_d&=148.02^{+3.63}_{-3.60}~\mathrm{Mpc}.
\end{align}
The Hubble constant $H_0$ and sound horizon $r_d$ are consistent with the results of Planck 2018.
However, there is a 2.4 $\sigma$ tension on $H_0$ with SH0ES 2019.
We provide the constraint on the Hubble constant in these two model-independent reconstructions,
and the mean values are a little larger than Planck 2018.
Meanwhile, the sound horizon $r_d$ is nicely consistent with the Planck results as well.
In the $\Lambda$CDM model and the two reconstructions of $H(z)$,
the results of $r_d$ are basically the same including the mean and the uncertainty.
We can conclude that the sound horizon $r_d$ is robust for the different parametrizations.
In other words, it is nearly free from dependence on the expansion history $H(z)$.

\section{Summary and conclusions}
\label{sec:sum}

In this paper, we provide a new independent measurement on the Hubble constant using the low-redshift observational data sets
including the measurements of BAO, observational $H(z)$ data, GW170817 and SN measurement.
In order to avoid imposing a prior of sound horizon $r_d$ from CMB measurement,
we remove the prior from Planck and set $r_d$ as a free sampling parameter in BAO distance measure, and we find
$H_0=68.63^{+1.75}_{-1.77}$ km s$^{-1}$ Mpc$^{-1}$, $r_d=146.85^{+3.29}_{-3.77}$ Mpc in $\Lambda$CDM model,
$H_0=68.02\pm1.82$ km s$^{-1}$ Mpc$^{-1}$, $r_d=148.18^{+3.36}_{-3.78}$ Mpc in the cubic expansion of $H(z)$,
and $H_0=68.58\pm1.76$ km s$^{-1}$ Mpc$^{-1}$, $r_d=148.02^{+3.63}_{-3.60}$ Mpc in the polynomial expansion of $H(z)$.
Our results of Hubble parameter $H_0$ and sound horizon $r_d$ are basically consistent with Planck 2018 in 1 $\sigma$.
However, $H_0$ is still in 2.4 $\sigma$, 2.6 $\sigma$, and 2.4 $\sigma$ tension with SH0ES 2019, respectively.
In addition, we conclude that $H_0$ and $r_d$ are both insensitive to the reconstruction of the expansion history.

\section*{Acknowledgments}

We acknowledge the use of HPC Cluster of ITP-CAS.
This work was supported by grants from NSFC (Grants No. 12005183, No. 11991052, No. 11975019, No. 11690021, No. 11947302), the Strategic Priority Research Program of Chinese Academy of Sciences (Grants No. XDB23000000, No. XDA15020701), Key Research Program of Frontier Sciences, CAS, Grant No. ZDBS-LY-7009, and the Natural Science Foundation of the Jiangsu Higher Education Institutions of China (Grant No. 20KJD140002).

\end{document}